\documentclass[journal]{IEEEtran}%

\usepackage{cite}

\ifCLASSINFOpdf
\usepackage[pdftex]{graphicx}
\else
\usepackage[dvips]{graphicx}
\fi

\usepackage[cmex10]{amsmath}
\usepackage{multirow}
\usepackage{chngpage}
\usepackage{graphicx, subfigure}
\usepackage[usenames]{color}
\usepackage[table]{xcolor}
\usepackage{makecell}
\usepackage{pgfplots}
\usepackage{tikz}
\usepackage{booktabs,siunitx}

\definecolor{lightgray}{gray}{0.9}
\definecolor{viola}{rgb}{0.7,0.1,1.0}
\definecolor{red}{rgb}{1.0,0.0,0.2}
\definecolor{comm}{gray}{0.0}
\definecolor{DarkGreen}{rgb}{0.000000,0.892157,0.500000}

\usepackage{algorithmic}
\usepackage{algorithm}
\usepackage[cmex10]{amsmath}
\usepackage{amsfonts}  
\usepackage{amssymb}
\usepackage{multirow}
\usetikzlibrary{decorations.pathreplacing,calc}

\def\hlinewd#1{%
\noalign{\ifnum0=`}\fi\hrule \@height #1 %
\futurelet\reserved@a\@xhline}

\newtheorem{theorem}{Theorem}[section]

\newenvironment{proof}[1][Proof]{\begin{trivlist}
\item[\hskip \labelsep {\bfseries #1}]}{\end{trivlist}}

\newcommand{\qed}{\nobreak \ifvmode \relax \else
      \ifdim\lastskip<1.5em \hskip-\lastskip
      \hskip1.5em plus0em minus0.5em \fi \nobreak
      \vrule height0.75em width0.5em depth0.25em\fi}

\IEEEoverridecommandlockouts

\begin{document}
%

%

	\title{A Sound and Complete Axiomatization of Majority-$n$ Logic}

\author{\IEEEauthorblockN{Luca Amar\'{u}, {\em Student Member, IEEE}, Pierre-Emmanuel Gaillardon, {\em Member, IEEE}, \\
Anupam Chattopadhyay, {\em Senior Member, IEEE}, Giovanni De Micheli, {\em Fellow, IEEE}}
\thanks{\footnotesize Luca Amar\'{u}, Pierre-Emmanuel Gaillardon and Giovanni De Micheli are with the Integrated Systems Laboratory, Swiss Federal Institute of Technology, Lausanne, EPFL, 1015 Lausanne, Switzerland (e-mail: luca.amaru@epfl.ch; pierre-emmanuel.gaillardon@epfl.ch; giovanni.demicheli@epfl.ch). 

Anupam Chattopadhyay is with Nanyang Technological University, 639798, Singapore (e-mail: anupam@ntu.edu.sg). 

}
}

\maketitle

\begin{abstract} \hspace{0.0001in}
Manipulating logic functions via majority operators recently drew the attention of researchers in computer science. 
For example, circuit optimization based on majority operators enables superior results as compared to traditional synthesis tools.  
Also, the Boolean satisfiability problem finds new solution approaches when described in terms of majority decisions. 
To support computer logic applications based on majority, a sound and complete set of axioms is required. 
Most of the recent advances in majority logic deal only with ternary majority (MAJ-3) operators because the axiomatization with solely MAJ-3 and complementation operators is well understood. 
However, it is of interest 
extending such axiomatization to $n$-ary majority operators (MAJ-$n$) from both the theoretical and practical perspective. 
In this work, we address this issue by introducing a sound and complete axiomatization of MAJ-$n$ logic. 
Our axiomatization naturally includes existing MAJ-3 and MAJ-5 axiomatic systems. Based on this general set of axioms, computer applications 
can now fully exploit the expressive power of majority logic. 

\vspace{0.15in}
\end{abstract}

\vspace{0.15in}

\section{Introduction}
\label{intro}

\IEEEPARstart{B}{oolean} logic and its axiomatization is fundamental to the whole field of computer science.
Traditionally, Boolean logic is axiomatized in terms of conjunction (AND), disjunction (OR) and complementation (INV) operators. 
Virtually, all of today's digital computation is performed
 by using these operators with their associated laws. Recently, it
was shown that more efficient logic 
computation is possible by using a majority operator in place of conjunction and disjunction operators\cite{Amaru, Amaru2, AmaruTCAD, IWLS}. 
Moreover, the properties of majority operators, such as stability, have been proved to be the best fit for solving important problems in computer science \cite{stablest,maj1,Sasao, SynATPG}. 
Regarding emerging technologies, 
majority operators are the natural 
logic primitives for several beyond-CMOS candidates \cite{fan2014_spin_mem, augustin_maj, datta_asl, imre_qca, snider99_qca, navi_maj5, qcadesigner, gao13_cmtl, hao14_dwm, li_dna, perkowski_maj_rev, coolswd, QCATOC, QCATOC2, ZhangTCAD}. 
In order to exploit the unique opportunity led by majority in computer applications, a sound and complete set of manipulation rules is required. Most of the recent studies on majority logic based computation
consider ternary majority (MAJ-3) operators because the axiomatization in this context is well understood. 
To unlock the real expressive power of majority logic, it is of interest to 
extend such axiomatization to $n$-ary ($n$ odd) majority operators (MAJ-$n$).

We introduce in this paper a sound and complete axiomatization of MAJ-$n$ logic. 
Our axiomatization is the natural extension of existing majority logic systems with fixed number of inputs. Based on the majority axioms
introduced in this work, computing systems 
can use at its best the expressive power of majority logic. 

The remainder of this paper is organized as follows. 
Section \ref{back} gives background and notations useful for
the rest of this paper.
Section \ref{axiom} introduces our sound and complete axiomatization for MAJ-$n$ logic. Section \ref{disc} discusses relevant applications of our majority logic system
in  logic optimization, Boolean satisfiability, repetition codes 
and emerging technologies. 
Section \ref{concl} concludes the paper.

\vspace{0.15in}

\section{Background and Notations}
\label{back}

We provide hereafter terms and notions
useful in the rest of the paper. We start by 
introducing basic notation and symbols for logic operators and we continue by presenting special properties of Boolean functions.
We define a compact vector notation for Boolean variables and discuss Boolean algebras with a particular emphasis on MAJ-3/INV Boolean algebra. 

\vspace{0.1in}

\subsection{Notations}
 In the binary Boolean domain, the symbol $\mathbb{B}$ indicates the set of binary values $\{0,1\}$;
the symbols $\land$ and $\lor$ represent the conjunction (AND) and disjunction (OR) operators; the symbol $\neg$ represents the 
complementation (INV) operator; and \texttt{0}/\texttt{1} represent the false/true logic values.
Alternative symbols for $\land$, $\lor$ and $\neg$ are $\cdotp$, $+$, and $'$, respectively. 

\vspace{0.1in}

\subsection{Self-Dual Function}
A logic function $f(x,y,..,z)$ is said to be {\em self-dual} if $f(x,y,..,z)=\neg f(\neg x,\neg y,..,\neg z)$ \cite{Sasao}. By complementation, an equivalent {\em self-dual}  formulation is 
$\neg f(x,y,..,z)=f(\neg x,\neg y,..,\neg z)$.

\vspace{0.1in}

\subsection{Majority Function}
An $n$-input ($n$ being odd) majority function $M_n$ is defined on reaching a threshold 
$\lceil n/2 \rceil$ of true inputs \cite{Sasao}. For example, the 
three input majority function $M_3(x,y,z)$ can be expressed as $\land,\lor$ by $(x\land y)\lor(x \land z)\lor(y \land z)$. Also 
$(x\lor y)\land(x \lor z)\land(y \lor z)$ is a valid representation for $M_3(x,y,z)$. The majority function is {\em self-dual}  \cite{Sasao}.
Note that an $M_n$ operator filled with $\lfloor n/2 \rfloor$  \texttt{0}/\texttt{1} collapses into a AND/OR operator \cite{Sasao}.

\vspace{0.1in}

\subsection{Vector Notation for Boolean Variables}
For the sake of compactness, we denote a container (vector) of $n-m+1$ Boolean variables 
by $x_m^n$, where the notation starts from index $m$ and ends at index $n$. 
When the actual length of the vector is not important, a
simpler notation for $x_m^n$ 
is boldface {\boldmath{$x$}}.
The element at index $i$ in vector $x_m^n$ is denoted by 
$x_i$. The complementation of a vector $x_m^n$ is denoted by $\neg x_m^n$ which means $\neg x_i$ $\forall i \in [m,m+1,..,n-1,n]$.
With this notation, the aforementioned self-dual property becomes $\neg f(x_m^n)=f(\neg x_m^n)$.
For the sake of clarity, we give an example about the vector notation. Let $(a,b,c,d,e)$ be 5 Boolean variables to be represented in vector notation. 
Here, the start/end indeces are $m=1$~/~$n=5$, respectively, and the vector itself is $x_1^5$. 
The elements of $x_1^5$ are $x_1=a$, $x_2=b$, $x_3=c$, $x_4=d$ and $x_5=e$.

\vspace{0.1in}

\subsection{Boolean Algebra}
\label{boolalg}
 The standard binary Boolean algebra (originally axiomatized by Huntington \cite{Huntington}) is a non-empty set $(\mathbb{B},\land,\lor,\neg,0,1)$ 
 subject to {\em 
identity, commutativity, distributivity, associativity, and complement} axioms over  
$\land,\lor$ and $\neg$\cite{Sasao,Brown}. For the sake of completeness, we report these basic axioms in Eq.~\ref{traditional}. 

\vspace{0.1in}

\begin{equation}
\label{traditional}
{\text{\large \boldmath{$\Delta$}}}\left\{\begin{array}{l l}
    \text{\bf Identity : \boldmath{$\Delta.I$}} \\x\lor 0=x\\ x\land 1=x\\
    \text{\bf Commutativity : \boldmath{$\Delta.C$}} \\ x\land y=y\land x\\ x\lor y=y\lor x \\
    \text{\bf Distributivity : \boldmath{$\Delta.D$}} \\ x \lor (y\land z) = (x\lor y)\land(x \lor z)\\ x \land (y \lor z)=(x \land y)\lor(x \land z)  \\
    \text{\bf Associativity : \boldmath{$\Delta.A$}} \\ x \land (y\land z) = (x\land y)\land z\\ x \lor (y \lor z)=(x \lor y)\lor z  \\
    \text{\bf Complement : \boldmath{$\Delta.Co$}} \\ x \lor \neg x =1\\ x\land \neg x =0 \\
  \end{array} \right.
\end{equation}

\vspace{0.2in}

This axiomatization for Boolean algebra is sound and complete \cite{Jonsson,Brown}. Informally, it means that, logic arguments or formulas, proved by axioms in $\Delta$ are valid (soundness) and 
all true logic arguments are provable (completeness). More precisely, it means that, in the induced logic system, all theorems are tautologies (soundness) 
and all tautologies are theorems (completeness). We refer the reader to \cite{Jonsson} for a more formal discussion on mathematical logic.  
In computer logic applications, only sound axiomatizations are of interest \cite{Brown}.  Complete and sound axiomatizations are desirable \cite{Brown}.

Other Boolean algebras exist, with different operators and axiomatizations, such as Robbins algebra,
Freges algebra,  Nicods algebra, MAJ-3/INV algebra, etc. \cite{Jonsson}. In the immediate following, we give details on the MAJ-3/INV Boolean algebra. 

\vspace{0.1in}

\subsection{MAJ-3/INV Boolean Algebra}
\label{MIGalg}
The MAJ-3/INV Boolean algebra introduced in \cite{Amaru} is
defined over the set $(\mathbb{B},M_3,\neg,0,1)$, where $M_3$ is the ternary majority operator 
and $\neg$ is the unary complementation operator.
The following set of five primitive transformation rules, referred to as $\Omega_3$, is an 
 {\em axiomatic system} for $(\mathbb{B},M_3,\neg,0,1)$. {\color{comm}All variables 
 belong to $\mathbb{B}$.}
 
 \vspace{0.1in}
 
\begin{equation}
\label{omega}
{\text{\large \boldmath{$\Omega_3$}}}\left\{\begin{array}{l l}
    \text{\bf Commutativity : \boldmath{$\Omega_3.C$}} \\M_3(x,y,z)=M_3(y,x,z)=M_3(z,y,x)\\
    \text{\bf Majority : \boldmath{$\Omega_3.M$}} \\ \left\{\begin{array}{l l}
    \text{if($x=y$): }M_3(x,y,z)=x=y\\ 
    \text{if($x=\neg y$): }M_3(x,y,z)=z\\ 
  \end{array} \right. \\
    \text{\bf Associativity : \boldmath{$\Omega_3.A$}} \\M_3(x,u,M_3(y,u,z))=M_3(z,u,M_3(y,u,x))\\
    \text{\bf Distributivity : \boldmath{$\Omega_3.D$}} \\M_3(x,y,M_3(u,v,z))=\\M_3(M_3(x,y,u),M_3(x,y,v),z)\\
    \text{\bf Inverter Propagation : \boldmath{$\Omega_3.I$}} \\\neg M_3(x,y,z)=M_3(\neg x,\neg y,\neg z)\\
  \end{array} \right.
\end{equation}

\vspace{0.15in}

It has been shown that this axiomatization is sound and complete with respect to $(\mathbb{B},M_3,\neg,0,1)$ \cite{Amaru}.
The MAJ-3/INV Boolean algebra finds
application 
in circuit optimization and has already showed some promising results \cite{Amaru}. 

Note that early attempts to majority logic 
have already been reported in the 60's \cite{akers_maj_manip,cohn_maj_axiom,lindaman_maj_network,Miller,Tohma,Miyata} 
but
they mostly focused on three input majority operators.
Also, derived
logic manipulation methods
failed to gain momentum due to their inherent complexity.

While traditional Boolean algebras can be naturally extended from 2 to $n$ variables, it is currently unclear how such a majority axiomatization extends to
an arbitrary number of variables $n$ (odd). In the following, we address this question by proposing a natural axiomatization of MAJ-$n$/INV logic. 

\section{Axiomatization of MAJ-$n$ Logic}
\label{axiom}

In this section, we present the generic axiomatization 
of MAJ-$n$ logic. We first extend the set of five axioms presented in \cite{Amaru} to $n$-variables, with $n$ being an odd integer.
Then, we show their validity in the Boolean domain. Finally, 
we demonstrate their completeness by inclusion of 
other complete Boolean axiomatizations. 

\subsection{Generic MAJ-$n$/INV Axioms} 
 
 The five axioms for MAJ-3/INV logic in \cite{Amaru}
 deal with {\em commutativity}, {\em majority},
 {\em associativity},
 {\em distributivity}, and
 {\em inverter propagation} laws. The following
 set of equations extends their domain to an arbitrary odd number
 $n$ of variables.
 Note that all axioms, hold with $n\ge 3$.
 
 \vspace{0.1in}
 
\begin{equation}
\label{omegan}
{\text{\large \boldmath{$\Omega_n$}}}\left\{\begin{array}{l l}
    \text{\bf Commutativity : \boldmath{$\Omega_n.C$}}\\M_n(x_1^{i-1},x_i,x_{i+1}^{j-1},x_j,x_{j+1}^{n})=\\M_n(x_1^{i-1},x_j,x_{i+1}^{j-1},x_i,x_{j+1}^{n})\\
    \text{\bf Majority : \boldmath{$\Omega_n.M$}} \\ 
    \text{If($\lceil \frac{n}{2} \rceil$ elements of $x_1^n$ are equal to $y$): }\\\hspace{0.1in} M_n(x_1^n)=y\\ 
    \text{If($x_i\neq x_j$): }\\\hspace{0.1in} M_n(x_1^n)=M_{n-2}(y_1^{n-2})\\\text{\hspace{0.1in} where $y_1^{n-2} = x_1^n$ removing $\{x_i,x_j\}$}\\
  
    \text{\bf Associativity : \boldmath{$\Omega_n.A$}} \\M_n(z_1^{n-2}, y, M_n(z_1^{n-2}, x,w)) =\vspace{0.1in}\\
M_n(z_1^{n-2}, x, M_n(z_1^{n-2}, y,w))\\

    \text{\bf Distributivity : \boldmath{$\Omega_n.D$}} \\M_n(x_1^{n-1},M_n(y_1^{n}))=\vspace{0.1in}\\M_n(M_n(x_1^{n-1},y_1),M_n(x_1^{n-1},y_2),...,\\\hspace{0.1in}M_n(x_1^{n-1},y_{\lceil \frac{n}{2} \rceil}),y_{\lceil \frac{n}{2} \rceil+1},...,y_n)=\vspace{0.1in}\\
 M_n(M_n(x_1^{n-1},y_1),M_n(x_1^{n-1},y_2),...,\\\hspace{0.1in} M_n(x_1^{n-1},y_{\lceil \frac{n}{2} \rceil+1}),y_{\lceil \frac{n}{2} \rceil+2},...,y_n)=     \vspace{0.1in}\\
 M_n(M_n(x_1^{n-1},y_1),M_n(x_1^{n-1},y_2),...,\\\hspace{0.1in} M_n(x_1^{n-1},y_{n-1}),y_n)\\
    \text{\bf Inverter Propagation : \boldmath{$\Omega_n.I$}} \\\neg M_n(x_1^n)=M_n(\neg x_1^n)\\
  \end{array} \right.
\end{equation}

\vspace{0.1in}

Commutativity means that changing the order of the variables in
$M_n$ does not change the result.
Majority defines a logic decision threshold (over $n \ge 3$ variables) and a 
hierarchical
reduction of majority operators with complementary variables. Note that $M_3(x,y,\neg y)=x$ as boundary 
condition.
Associativity says that swapping pairs of variables
between cascaded $M_n$ sharing $n-2$
variables does not change the result. In this context, it is
important to recall 
that $n-2$ is an odd number if $n$ is an odd number.
Distributivity delimits the re-arrangement freedom of variables
over cascaded $M_n$ operators.
Inverter propagation moves complementation freely 
from the outputs to the inputs of a $M_n$ operator, and 
{\em viceversa}.

For the sake of clarity, we give an example for each axiom over a finite $n$-arity. 

Commutativity with $n=5$: \\$M_5(a,b,c,d,e)=M_5(b,a,c,d,e)=M_5(a,b,c,e,d)$.

Majority with $n=7$: \\$M_7(a,b,c,d,e,g,g')=M_5(a,b,c,d,e)$.

Associativity with $n=5$: \\$M_5(a,b,c,d,M_5(a,b,c,g,h))=M_5(a,b,c,g,M_5(a,b,c,d,h))$.

Distributivity with $n=7$: \\$M_7(a,b,c,d,e,g,M_7(x,y,z,w,k,t,v))=M_7(M_7(a,b,c,d,e,g,x),M_7(a,b,c,d,e,g,y),\\ \text{\hspace{0.1in}}M_7(a,b,c,d,e,g,z),M_7(a,b,c,d,e,g,w),k,t,v)$.

Inverter propagation with $n=9$: \\$\neg M_9(a,b,c,d,e,g,h,x,y)=M_9(\neg a,\neg b,\neg c,\neg d,\neg e,\neg g,\neg h,\neg x,\neg y)$.

\subsection{Soundness}

To demonstrate the validity of these laws, and thus the validity
of the MAJ-$n$ axiomatization, we need to show that each equation in $\Omega_n$ is sound with respect to the original domain, i.e., $(\mathbb{B},M_n,\neg,0,1)$
\footnote{By $M_n$, it is intended 
any $M_i$ with $i \le n$. Indeed, 
any $M_i$ operator with $i\le n$ can be emulated by a fully-fed $M_n$ operator with pairs of regular/complemented variables, e.g., $M_5(a,b,c,d,\neg d)=M_3(a,b,c).$}.
The following theorem addresses this requirement.

\vspace{0.05in}
\begin{theorem}
\label{sound}
Each axiom in $\Omega_n$ is sound (valid) w.r.t. $(\mathbb{B},M_n,\neg,0,1)$.
\end{theorem}
\vspace{0.05in}
\begin{proof}

{\bf Commutativity \boldmath{$\Omega_n.C$}} Since majority is defined on reaching a threshold 
$\lceil n/2 \rceil$ of true inputs then it is independent of the order of its inputs. This means 
 that changing the order of
operands in $M_n$ does not change the output value.
Thus, this axioms is valid in $(\mathbb{B},M_n,\neg,0,1)$.

{\bf Majority \boldmath{$\Omega_n.M$}} Majority 
first defines
the output behavior of $M_n$ in the Boolean domain. Being a
definition, it does not need particular proof for soundness.
Consider then the second part of the majority axiom.
The recursive inclusion of $M_{n-2}$ derives from the mutual
cancellation of complementary variables. In a binary majority voting system of $n$ electors, two electors voting 
to opposite values annihilate themselves. The final decision
is then just depending on the votes from the remaining $n-2$
electors. Therefore, this axiom is valid in $(\mathbb{B},M_n,\neg,0,1)$.

{\bf Associativity \boldmath{$\Omega_n.A$}} We split this proof in three parts that cover the whole Boolean space. Thus, it is sufficient to prove the validity of the
associativity axiom for each of these parts. {\bf (1) the vector $z_1^{n-2}$ contains at least one logic 1 and one logic 0.} In this case, it is possible to apply $\Omega_n.M$ and reduce
$M_n$ to $M_{n-2}$. If we remain in case (1), we can keep applying $\Omega_n.M$. At some point, we will end up in case (2) or (3).  {\bf (2) the vector $z_1^{n-2}$ contains all logic 1.} 
For $n>3$, the final voting decision is 1 for both equations, so the equality holds. In case $n=3$
and the the vector $z_1^{n-2}$ contains all logic 1, the majority operator collapses into a disjunction operator.
For example, $M_3(1,a,M_3(1,c,d))=\lor_2(a,\lor_2(c,d))$. 
 Here, the validity of the associativity axiom follows then 
from traditional disjunction associativity. {\bf (3)
the vector $z_1^{n-2}$ contains all logic 0.} For $n>3$, the final voting decision is 0 for both equations, so the equality holds. In case $n=3$
and the vector $z_1^{n-2}$ contains all logic 0, the majority operator collapses into a conjunction operator.
For example, $M_3(0,a,M_3(0,c,d))=\land_2(a,\land_2(c,d))$.  Here, the validity of the associativity axiom follows then 
from traditional conjunction associativity. 

{\bf Distributivity \boldmath{$\Omega_n.D$}} We split this proof in three parts that cover the whole Boolean space. Thus, it is sufficient to prove the validity of the
distributivity axiom for each of these parts. 
Note that the distributivity axiom deals with a majority operator $M_n$ where one inner variable is actually another independent majority operator $M_n$.
Distributivity rearranges the computation in $M_n$ moving up the variables at the bottom level and down the variables at the top level. 
In this part of the proof we show that such rearrangement does not change the 
functionality of $M_n$, i.e., the final voting decision in $\Omega_n.D$.
Recall that $n$ is an odd integer greater than $1$ so $n-1$ must be an even integer. 
{\bf (1) half of $x_1^{n-1}$ values are logic 0 and the remaining half are logic 1.} In this case, the final voting decision in 
axiom $\Omega_n.D$ only depends on $y_1^n$. Indeed, all elements in 
$x_1^{n-1}$ annihilate due to axiom $\Omega_n.M$. In the two
identities of $\Omega_n.D$, we see that when $x_1^{n-1}$ annihilate
the equations simplify to $M_n(y_1^n)$, according to the predicted behavior.
{\bf (2) at least $\lceil n/2 \rceil$ of $x_1^{n-1}$ values are logic 0.} 
Owing to $\Omega_n.M$, the final voting decision in this case 
is logic 0. This is because more than half of the variables are logic 0
matching the prefixed voting threshold. In the two identities of $\Omega_n.D$, we see that more than half of the inner $M_n$ evaluate to logic 0 by direct application
of $\Omega_n.M$. In the subsequent phase, also the outer $M_n$ evaluates
to logic 0, as more than half of the variables are logic 0, according to the predicted behavior.
{\bf (3) at least $\lceil n/2 \rceil$ of $x_1^{n-1}$ values are logic 1.} This case
is symmetric to the previous one. 

{\bf Inverter Propagation \boldmath{$\Omega_n.I$}} Inverter propagation moves complementation from output to inputs, and {\em viceversa}. 
This axiom is a special case of the self-duality property previously presented. It holds for all majority operators in 
$(\mathbb{B},M_n,\neg,0,1)$.

\end{proof}
\vspace{0.05in}

The soundness of $\Omega_n$ in $(\mathbb{B},M_n,\neg,0,1)$ guarantees that 
repeatedly applying  $\Omega_n$ axioms to a Boolean formula we do not 
corrupt its original functionality. This property is of 
interest in logic manipulation systems where functional correctness is 
an absolute requirement. 

\subsection{Completeness}

While soundness speaks of the correctness of a logic systems, 
completeness speaks of its manipulation capabilities. 
For an axiomatization to be complete, all possible 
manipulations of a Boolean formula must be attainable by a sequence, possibly long, 
of primitive axioms.

We study the completeness of $\Omega_n$ axiomatization by comparison
to other complete axiomatizations of Boolean logic. The following theorem 
shows our main result.

\vspace{0.05in}
\begin{theorem}
\label{complete}
The set of five axioms in $\Omega_n$ is complete w.r.t. $(\mathbb{B},M_n,\neg,0,1)$.
\end{theorem}
\vspace{0.05in}
\begin{proof}
We first consider $\Omega_3$ and we show that it is complete w.r.t. $(\mathbb{B},M_3,\neg,0,1)$.
We need to prove that every valid argument, i.e., 
$(\mathbb{B},M_3,\neg,0,1)$-formula, has a proof in the system $\Omega_3$. By contradiction, suppose that a true $(\mathbb{B},M_3,\neg,0,1)$-formula, say $\alpha$, cannot 
be proven true using $\Omega_3$ rules. Such $(\mathbb{B},M_3,\neg,0,1)$-formula $\alpha$ can always be reduced into a $(\mathbb{B},\land,\lor,\neg,0,1)$-formula. Indeed, 
recall that $M(x,y,z)=(x\lor y)\land(x \lor z)\land(y \lor z)$.
Using $\Delta$, all $(\mathbb{B},\land,\lor,\neg,0,1)$-formulas can be proven, including $\alpha$. However, every $(\mathbb{B},\land,\lor,\neg,0,1)$-formula is also contained by $(\mathbb{B},M_3,\neg,0,1)$,
where $\land$ and $\lor$ are emulated by majority operators. Moreover, rules in $\Omega_3$ with one input fixed to $0$ and $1$ behaves as $\Delta$ rules (Eq. \ref{traditional}).
For example, $\Omega_3.A$ with variable $u$
fixed to logic 1 (0) behaves as $\Delta.A$ for disjunction
(conjunction). 
The other axioms follow analogously.
  This means that 
also $\Omega_3$ is capable to prove the reduced $(\mathbb{B},M,\neg,0,1)$-formula $\alpha$, contradicting our assumption. 
Thus $\Omega_3$ is complete w.r.t. $(\mathbb{B},M_3,\neg,0,1)$. 

We consider now  $\Omega_n$. First note that $(\mathbb{B},M_n,\neg,0,1)$ naturally includes 
$(\mathbb{B},M_3,\neg,0,1)$. Similarly, $\Omega_n$ axioms inherently 
extend the ones in $\Omega_3$. Thus, the completeness property is inherited 
provided that $\Omega_n$ axioms are sound. 
 However, $\Omega_n$ soundness is already proven in Theorem \ref{sound}. 
 Thus, $\Omega_n$ axiomatization is also complete. 
\end{proof}
\vspace{0.05in}

Being sound and complete, the axiomatization $\Omega_n$ defines a consistent framework to operate on 
Boolean logic via $n$-ary majority operators and inverters. 
In the following section, we discuss some promising 
applications in computer science of such majority logic system. 

\section{Discussion}
\label{disc}

In this section, we discuss relevant application of  $\Omega_n$
axiomatization. We first present the potential of logic optimization 
performed via MAJ-$n$ operators and inverters. Then, we show 
how Boolean satisfiability can be described in terms of majority 
operators
and solved using $\Omega_n$.
Successively, we demonstrate the manipulation of
repetition codes via $\Omega_n$ under a majority logic decoding 
scheme. 
Finally, we discuss the application of majority logic
to several emerging technologies, such as 
quantum-dot cellular automata, spin-wave devices, 
threshold logic and others.

\subsection{Logic Optimization}
\label{optimization}

Logic optimization is the process of manipulating a logic data
 structure, such as a logic circuit, 
 in order to minimize some target metric \cite{DeMicheli}. 
 Usual optimization targets are size (number of nodes/elements),
 depth (maximum number of levels) and interconnections (number of edges/nets). More elaborated 
 targets use a combination of size/depth/interconnections metrics, such as nodes$\times$interconnections and others.

Theoretical results from computer science show that 
majority logic 
circuits are much more compact than traditional ones based on conjunction and 
disjunction operators \cite{maj1}. 
For example, majority logic circuits of depth 2 and 3 possess the expressive power to represent arithmetic functions, such as powering, multiplication, division, addition etc., in polynomial size \cite{maj1}. On the other hand, the traditional AND/OR-based counterparts are exponentially sized \cite{maj1}. 

Given the existence of very compact majority logic circuits, we need an efficient set of manipulation laws to reach those circuits automatically. 
In this context, the axiomatic system previously introduced is the natural set of tools addressing this need. For example, consider a logic circuit (or Boolean function)
$f=M_5(M_3(a,b,c),M_3(a,b,d),M_3(a,b,e),M_3(a,b,g),h)$.
In circuit optimization, a common problem is to minimize the number of elements while keeping short some input-output paths. 
Suppose we want to minimize the number
of majority operators while keeping the path $h$ to $f$ as short as possible, i.e., one majority operator.  
The original circuit cost is 5 majority operators. 
To manipulate this formula, we first equalize 
the $n$-arity of the majority operators
using axiom $\Omega_n.M$, i.e., by adding a fake annihilated variable $x$, as:

$f=M_5(M_5(a,b,c,x,\neg x),M_5(a,b,d,x,\neg x),\\M_5(a,b,e,x,\neg x),M_5(a,b,g,x,\neg x),h)$

At this point, we can apply $\Omega_n.D$ and save one
majority operator as:

$f=M_5(M_5(a,b,c,x,\neg x),M_5(a,b,d,x,\neg x),\\M_5(a,b,e,x,\neg x),g,h)$.

Finally, we can reduce the majority $n$-arity to its minimum
via $\Omega_n.M$ as:

$f=M_5(M_3(a,b,c),M_3(a,b,d),M_3(a,b,e),g,h)$.

The resulting circuit cost is 4 majority operators.

\subsubsection{Optimization Script}

As emerged from the previous optimization example, 
an intuitive heuristic to optimize
majority logic circuits
consists of majority inflation rules (from $\Omega_n$)
followed by majority reduction rules (from $\Omega_n$). 
Alg.~\ref{topcontrol} depicts a simple
optimization script and a brief description follows. 
\begin{algorithm}[!ht]

\textbf{INPUT:} Majority Logic Network. \hspace{0.8in}\\ \textbf{OUTPUT:} Optimized Majority Logic Network.

\caption{Majority Logic Optimization Heuristic}
\begin{algorithmic}
\STATE{Majority Operator Increase n-arity($\Omega_n.M$);{\color{blue}\\// increase n-arity of the majority operator}}
\STATE{Majority Operator Simplifcation($\Omega_n.A, \Omega_n.D, \Omega_n.M$);{\color{blue}\\// deleting redundant majority operators}}
\STATE{Majority Operator Reduce n-arity($\Omega_n.M$);{\color{blue}\\// decrease n-arity of the majority operator}}
\end{algorithmic}
\label{topcontrol}
\end{algorithm}
First, the $n$-arity of all majority operators in the logic circuit is temporarily increased by using $\Omega_n.M$ rule from right to left, for example $M_3(a,b,c)=M_5(a,b,c,\neg c, c)$. This operation 
unlocks new simplification opportunities. 
Then, redundant majority operators are identified 
and deleted through $\Omega_n.A, \Omega_n.D, \Omega_n.M$ rules. 
Finally, the $n$-arity of all majority operators in the 
logic circuit
is decreased to the minimum via $\Omega_n.M$ rule
from left to right.

This approach naturally targets depth and size reductions 
in the majority logic network. However, 
it can be extended to target 
more elaborated metrics, such as $\sum_{i=1}^{M}  fanin(node_{i})$ or $M\times N_{inv}$, where $M$ is the total number of nodes and $N_{inv}$ is the number of inverters. The best metric 
depends on the considered technology for final
implementation.

\subsubsection{Full-Adder Case Study}

In order to prove the efficacy of the 
majority optimization heuristic in Alg.~\ref{topcontrol}, 
we consider as case study the full-adder logic circuit.
The full-adder logic circuit 
is fundamental to most arithmetic circuits.
Consequently, the effective optimization of full-adders is of
paramount importance. 

A full-adder represents a three-input and 
two-output Boolean function:

$sum= a\oplus b \oplus c_{in}$

$c_{out}=M_3(a,b,c)$

\begin{figure}[!ht]
\centering%
\includegraphics[width=0.5\columnwidth]{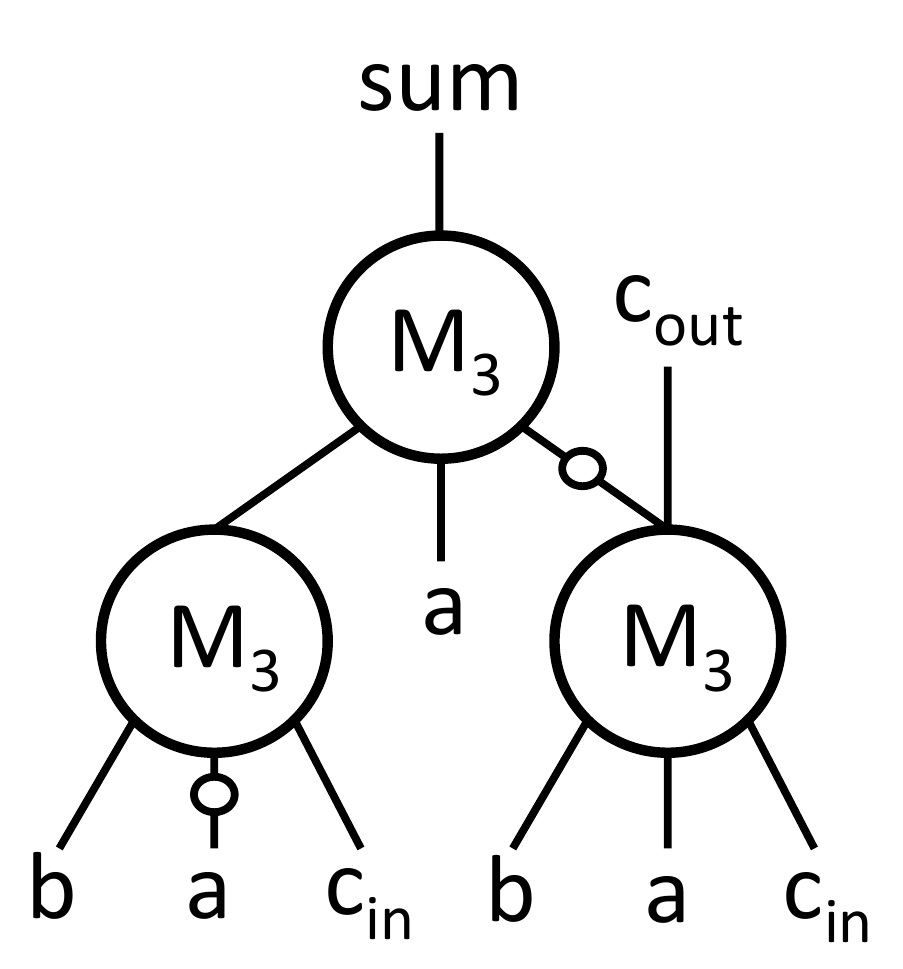}
\caption{Majority logic circuit for the full-adder with
operator $n$-arity equal to 3.
Complementation is represented by bubbles on the edges.}
\label{fam3}
\end{figure}

Using just majority operators with $n$-arity equal to three, 
the best full-adder implementation counts 3 majority
nodes, inverters apart, as depicted by Fig. \ref{fam3}. 
However, a more compact majority logic
network is possible by exploiting higher $n$-arity degrees
and manipulating such
majority logic circuit via $\Omega_n$.
In particular, the critical operation is $sum$ because
$c_{out}$ is naturally represented by a single
$M_3$ operator. So, for $sum$ our
optimization heuristic first expands the top 
majority operator from an $n$-arity of three

$sum=M_3(a,\neg M_3(a,b,c_{in}),M_3(\neg a,b,c_{in}))$

  to an $n$-arity 
of 5 as 

$sum=M_5(a,\neg M_3(a,b,c_{in}),\neg M_3(a,b,c_{in}),\\ M_3(a,b,c_{in}),M_3(\neg a,b,c_{in}))$.

After that, derived simplification rules from $\Omega_n$, 
called relevance rules in \cite{Amaru}, reduce the number
of majority operators to 2 as

$sum=M_5(a,\neg M_3(a,b,c_{in}),\neg M_3(a,b,c_{in}), b,c_{in})$.

\begin{figure}[!ht]
\centering%
\includegraphics[width=0.45\columnwidth]{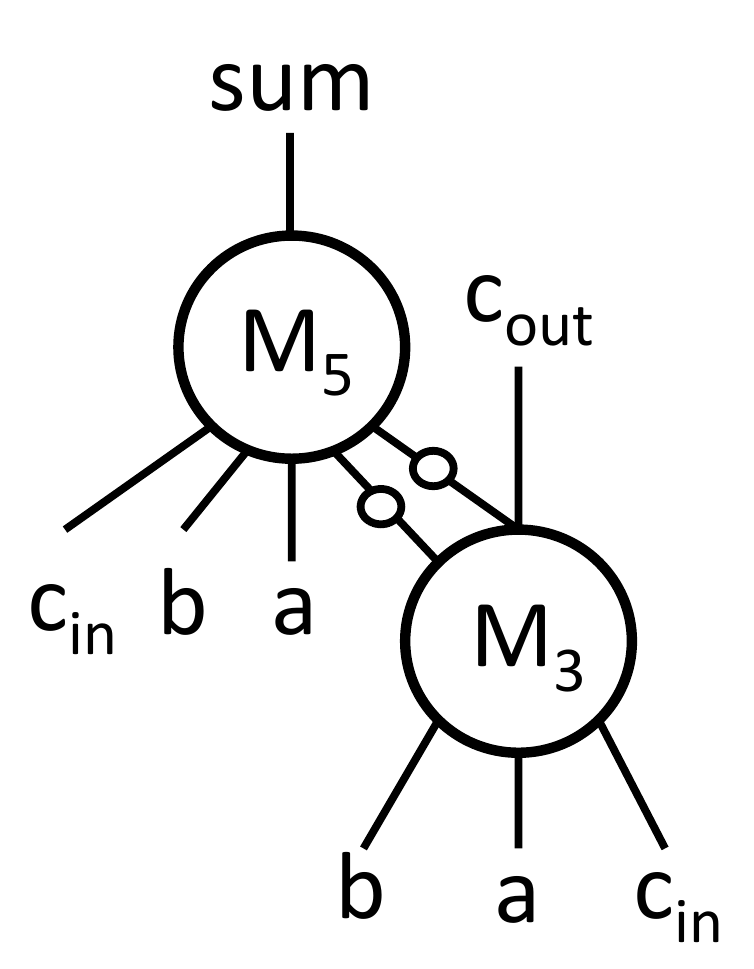}
\caption{Majority logic circuit for the full-adder with
unbounded 
operator $n$-arity.
Complementation is represented by bubbles on the edges.}
\label{fam5}
\end{figure}

In its graph representation, depicted by Fig.~\ref{fam5}, 
this representation of {\em sum} just consists of two majority operators as the internal $M_3(a,b,c_{in}),$ is shared.
Moreover, $M_3(a,b,c_{in})$ is also generating the 
$c_{out}$ function which can be further shared.
This means that the optimized logic circuit  in Fig.~\ref{fam5}, counting just two majority 
operators, is a minimal implementation for the full-adder
 in terms
of majority logic.
\begin{figure}[!ht]
\centering%
\includegraphics[width=0.67\columnwidth]{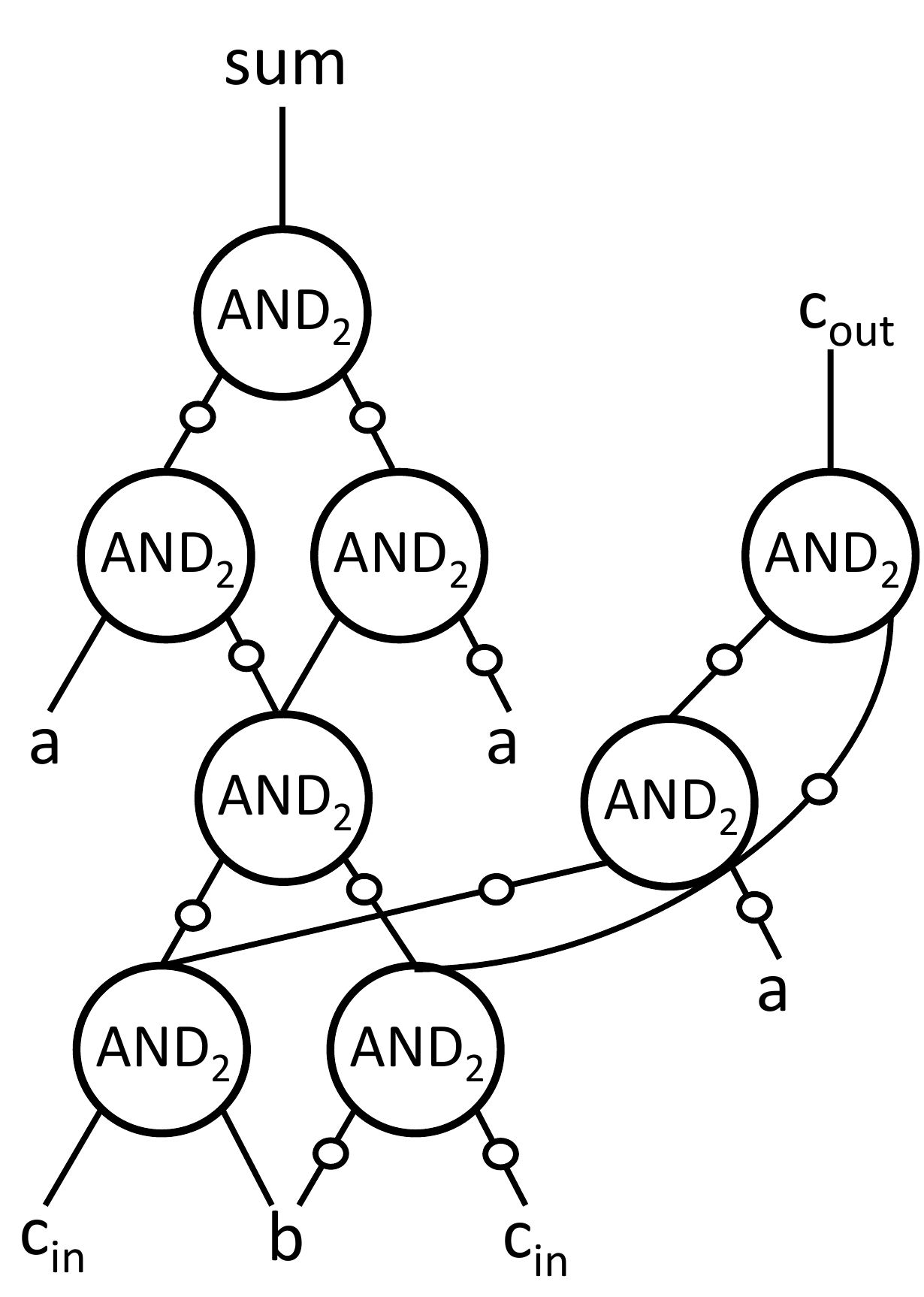}
\caption{AND-inverter logic circuit for the full-adder
optimized via ABC academic tool.
Complementation is represented by bubbles on the edges.}
\label{fan}
\end{figure}
To provide a reference, an optimized AND-inverter graph
representation for the full-adder is depicted by Fig.~\ref{fan}.
It counts 8 nodes and has been optimized using 
the state-of-the-art academic ABC optimizer 
\cite{ABC} which manipulates
AND-inverter graphs. We can see that the majority 
logic circuit produced by our optimization
heuristic is much more compact thanks to the majority 
logic expressiveness and 
to the properties of our axiomatic system, $\Omega_n$.

The minimality of the majority logic circuit in Fig. \ref{fam5} is
 formally proved in the following theorem. 

\vspace{0.05in}
\begin{theorem}
\label{faopt}
The majority logic circuit  in Fig.~\ref{fam5} for the full-adder
has the minimum number of majority operators.
\end{theorem}
\vspace{0.05in}
\begin{proof}
The full-adder consists of two distinct functions. Being distinct, they require at least two separate majority operators fed
with different signals. The majority logic circuit in
 Fig.~\ref{fam5}
actually consists of two majority operators
thus being minimal. 
\end{proof}
\vspace{0.05in}

On top of having the minimum number of operators, the 
 majority network  in Fig.~\ref{fam5} has lower 
$\sum_{i=1}^{M}  fanin(node_{i})$ metric (equal to 8) as compared to the majority network in Fig.~\ref{fam3} (equal to 9). The number of inverters is 2 in both cases.

We see that the axiomatic system $\Omega_n$ 
can be used to optimize majority logic circuits and 
produces excellent results. 
As the $\Omega_n$ rules 
are simple enough to be programmed on a computer, 
MAJ-$n$ logic optimization can be automated and applied to large systems.

\subsection{Boolean Satisfiability}

Boolean {\em satisfiability} (SAT) is the first known NP-complete
problem \cite{Garey}. Traditionally, SAT is 
formulated in  {\em Conjunctive Normal Form} (CNF)
\cite{Biere}. Recently, majority logic has been 
considered as an alternative to CNF to speed-up SAT \cite{IWLS}.
In \cite{IWLS}, a {\em Majority Normal Form} (MNF)
has been introduced, which is
a majority of majorities, 
where majorities are fed with literals, {\texttt 0} or {\texttt 1}.
The MNF-SAT problem is NP-complete in 
its most general definition \cite{IWLS}. 
However, there are interesting restrictions of MNF whose satisfiability can 
instead be decided in polynomial time. 
For example, when there are no mixed
logic constants appearing in the MNF, 
the MNF-SAT problem can be solved in polynomial time. 
This result is valid not just for MNF but for 
majority logic circuits in general \cite{IWLS}. 

In order to solve the general problem of majority logic 
satisfiability, and thus of MNF-SAT, a set
of manipulation rules is needed. Indeed, 
the core of most modern SAT solving tools 
make extensive use of Boolean logic axioms. 
When dealing with majority logic, our proposed
axiomatic system $\Omega_n$ is the natural tool to operate
on MNF forms, or alike, and prove their satisfiability.

\vspace{0.05in}

For the sake of clarity, we give an example
of majority SAT solving via $\Omega_n$ laws.
We consider not just an MNF, which is a two level 
logic representation form, but a general formula in 
$(\mathbb{B},M_n,\neg,0,1)$. 
Our example is the {\em unSAT} function
$f=M_5(M_3(a, b, c),M_5(M_5(a,b,c,0,0),\neg b, c,0,0),\neg a, \neg b, 0)$.
In oder to check the satisfiability of $f$, a majority SAT solver first tries to enforce at least 3 over 5 logic \texttt{1} in the top $M_5$ \cite{IWLS}. 
Otherwise, a conflict 
in the input assignment appears. If all possible input
assignments lead to a conflict the function is declared
unsatisfiable \cite{IWLS}. 

Let us first focus on the element 
$M_5(M_5(a,b,c,0,0),\neg b, c,0,0)$. Here, even before
 looking for possible assignments, our 
axiom $\Omega_n.A$  re-arranges 
the variables as $M_5(M_5(\neg b,b,c,0,0),a, c,0,0)$.
In this formula, our axiom
$\Omega_n.M$ directly annihilates $b$ and $\neg b$
leading to $M_5(M_3(c,0,0),a, c,0,0)$. Furthermore, 
$\Omega_n.M$ still applies twice corresponding to 
$M_5(0,a, c,0,0)$ and then $0$. We can substitute
this to the original formula
as 
$f=M_5(M_3(a, b, c),0,\neg a, \neg b, 0)$
  which symplifies the SAT problem. Now,
  we need both $\neg a$ and $\neg b$ to be {\texttt 1}
 in order to do avoid an immediate conflict.
This means
  $a=0$ and $b=0$. However, this assigment evaluates 
  always to {\texttt 0} the term 
  $M_3(a,b,c)$ generating a conflict for all
  input patterns. Thus, the original formula is declared
  unsatisfiable. 

As we can see, our majority logic axiomatic system 
$\Omega_n$
is the ground for proving the 
satisfiability of formula in $(\mathbb{B},M_n,\neg,0,1)$. 
Without $\Omega_n$, SAT tools would need to 
decompose all majority operators in AND/ORs 
because with conjunctions and disjunctions the classic set of 
Boolean manipulation rules apply. However, 
such decomposition would nullify the 
competitive advantage enabled by the 
majority logic
expressiveness. In this scenario, our $\Omega_n$
rules
fill the gap for manipulating majority operators natively.

\subsection{Decoding of Repetition Codes}

Repetition codes are basic error-correcting codes. The main
rationale in using repetition codes is to transmit a message several times over a noisy channel hoping that the channel corrupts only a minority of
the bits\cite{Massey}. In this scenario, decoding the received message
via majority logic is the natural way to correct transmission errors. 

Consider safety-critical communication systems. It
is common to have hierarchical levels of coding 
to decrease the chance of error and thus resulting in
system malfunction. When applied on several levels, 
majority logic decoding is nothing but a majority
logic circuit. The maximum number of cascaded
majority operators
determines the decoding performance.
We want to maximize the decoding performance 
while keeping the error probability low. In this scenario, 
we can use our axiomatic system $\Omega_n$
to explore different tradeoffs in depth/size manipulation 
of the corresponding majority decoding scheme. 

For the sake of clarity, we give an example
of the optimization for majority logic decoding via $\Omega_n$.
 Consider a 
 safety-critical communication system sending the same binary message
 $a$ over 5 different channels $C_1$, $C_2$, 
 $C_3$, $C_4$ and $C_5$. Each channel is affected
 by different levels of noise requiring just
 1 repetition for $C_1$, $C_2$, $C_3$, and $C_4$
but 5 repetitions for $C_5$.
Suppose also the communication over channel 5 is much
slower than in the other channels.
  The final decoded
 message is the majority of the each decoded message per channel. If we name $x_i$ the decoded message $a$
 for $i$-th channel  and $y$ the final decoded message,
 the system can be represented in majority logic
 as $y=M_5(x_1,x_2,x_3,x_4,x_5)$. 
Note that for $x_1$, $x_2$, $x_3$, $x_4$ the decoded message is actually identical to the received message because only 1 repetition is sent over the channels. 
 The element 
 $x_5$ is the only one needing further majority decoding, namely
 $x_5=M_5(z_1,z_2,z_3,z_4,z_5)$ where $z_i$ are the received $a$ messages over channel $C_5$. The final 
system is then expressable as $y=M_5(x_1,x_2,x_3,x_4,M_5(z_1,z_2,z_3,z_4,z_5))$.
 To decode the final message $y$,
 the critical element for 
 perfomance is $M_5(z_1,z_2,z_3,z_4,z_5)$, 
 with $z_5$ being the latest arriving message to 
 be processed. In this context, 
 we can use $\Omega_n.D$ axiom to redistribute
 the decoding operations and obtain an improvement
 in performance, which is not a trivial process. The idea is to push to the top 
 majority level $z_i$ variables, with the highest possible
 $i$ index.  For this purpose, 
 axioms  $\Omega_n.D$ transforms 
 $y=M_5(x_1,x_2,x_3,x_4,M_5(z_1,z_2,z_3,z_4,z_5))$
 into 
 $y=M_5(M_5(x_1,x_2,x_3,x_4,z_1),
 M_5(x_1,x_2,x_3,x_4,z_2),\\
 M_5(x_1,x_2,x_3,x_4,z_3),
 z_4,z_5)$.
 In this latter model of majority decoding,
 most of the computation is performed in advance
 before the late messages $z_4$ and $z_5$ 
 arrive. This means that, when the late $z_5$ arrives,
 there is need for just one level of majority computation
 and not two as in the initial model.

\subsection{Emerging Technologies}
Majority gates with more than $3$ inputs have been simulated and implemented for a variety of non-CMOS technologies. A further generalization of majority gates is threshold logic gate \cite{maj1}, which performs weighted sum of multiple inputs and once the sum is more than a pre-determined threshold, the output is true. As such, a threshold logic gate can be configured to function as a majority logic gate. In the following, we describe a few published works that describes majority or threshold gates with more than $3$ inputs. 

\begin{figure}[!ht]
\centering%
\includegraphics[width=1.05\columnwidth]{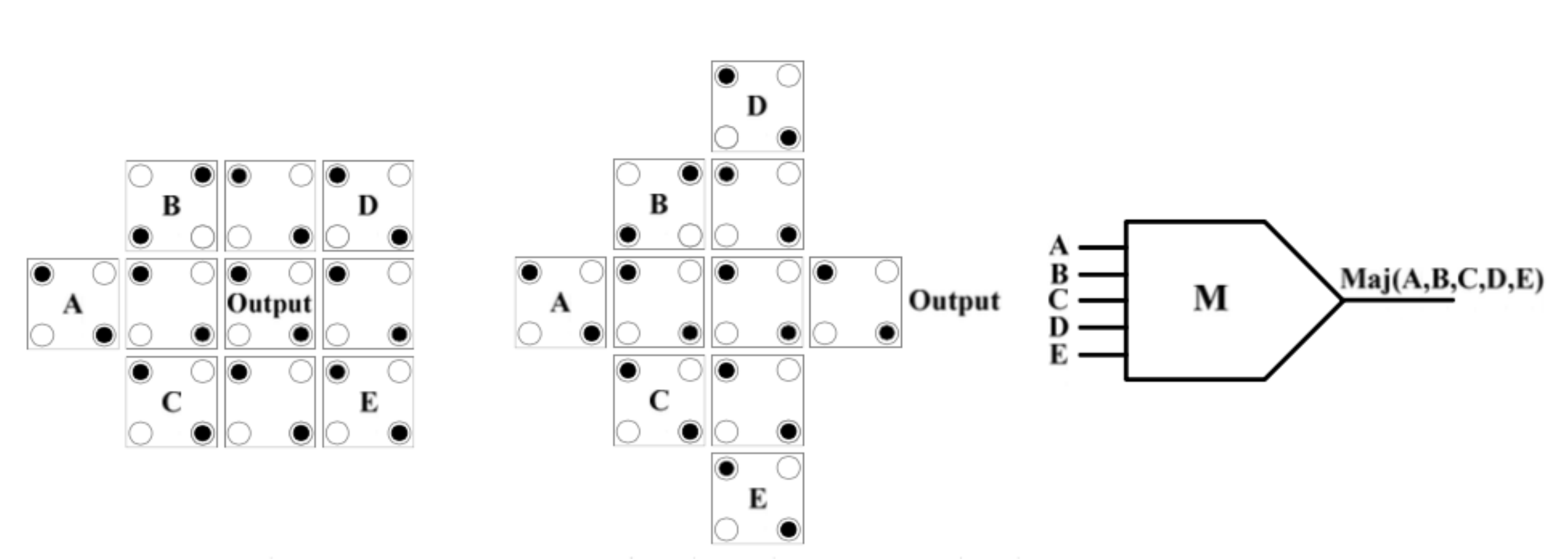}
\caption{Two different implementations of a
$M_5$ gate in QCA technology \cite{navi_maj5}.}
\label{maj5impl}
\end{figure}

\begin{figure}[!ht]
\centering%
\includegraphics[width=0.9\columnwidth]{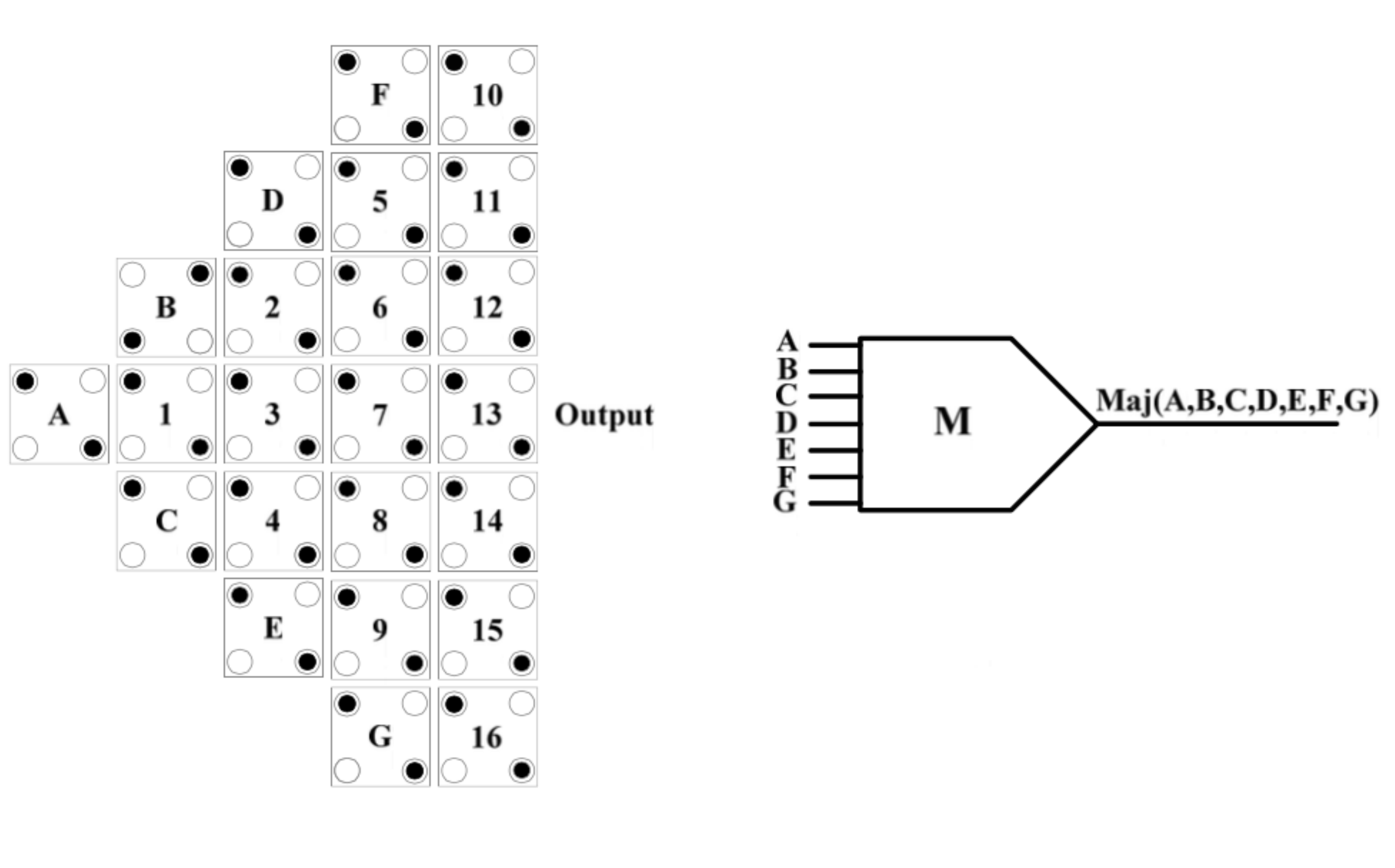}
\caption{Physical implementation of a
$M_7$ gate in QCA technology \cite{navi_maj5}.}
\label{maj7impl}
\end{figure}

Majority logic gates were experimentally demonstrated with {\em Quantum-dot Cellular Automata} (QCA) in~\cite{imre_qca} and~\cite{snider99_qca}. For facilitating QCA circuit design, a tool named QCADesigner is developed~\cite{qcadesigner}. Simulation of $M_5$ gate using QCADesigner is presented in several papers, including~\cite{navi_maj5}. 
Fig.~\ref{maj5impl} depicts two possible
QCA implementations for a $M_5$ gate.
Applications of large majority gates towards efficient adder construction were also discussed. For example, a $M_7$ has also been 
proposed.  
Fig.~\ref{maj7impl} depicts a possible
QCA implementation for a $M_5$ gate.

   Note that a $M_5$ gate, a $M_3$ gate 
and an inverter gate are sufficient to build a
full-adder, as highlighted 
by the
theoretical 
  case study in Section \ref{optimization}. 
In this 
scenario, the proposed $\Omega_n$ axiomatic 
system is key to unveil such efficient 
circuit implementations
in QCA nanotechnology, where majority gates 
are the logic primitives for computation. 

Very recently, 
a majority logic circuit based on domain-wall nanowires has been proposed in~\cite{hao14_dwm}. The circuit is used for computing binary additions efficiently and can be shown to scale for majority gates with arbitrary number of inputs.

\begin{figure}[!ht]
\centering%
\includegraphics[width=1.0\columnwidth]{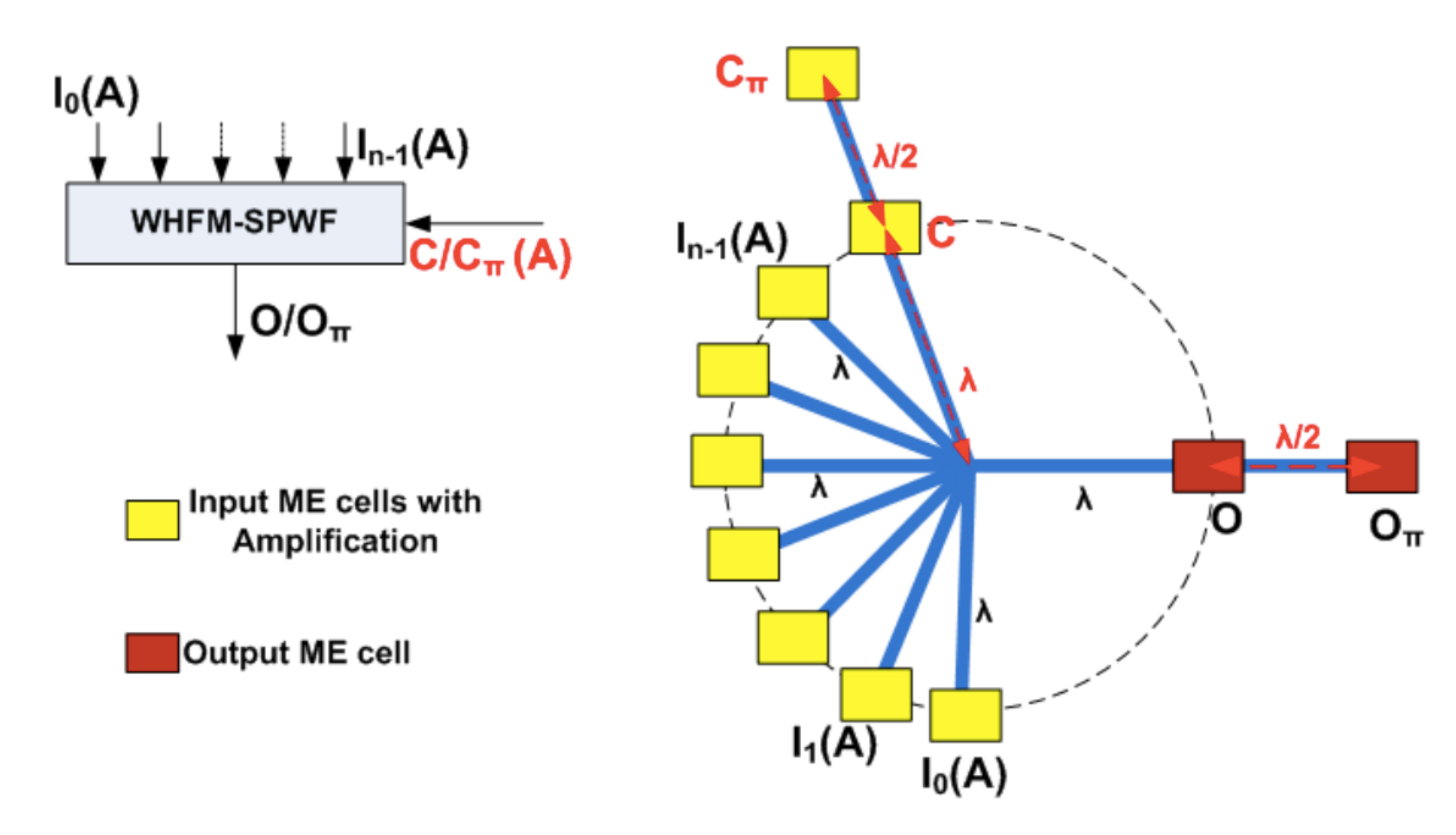}
\caption{
Block diagram and schematic representation of  a
high fan-in majority gate in spin-wave technology \cite{coolswd}.}
\label{cswd}
\end{figure}

All-spin logic gates are originally proposed in~\cite{datta_asl}. Majority logic gates using all-spin logic is proposed in~\cite{augustin_maj}. There, layout of $M_3$ gate using all-spin logic is shown and it is noted that majority gates with larger number of inputs can also be implemented. Indeed, 
a high fan-in majority gate is realizable by 
a simple
superposition of spin-waves with same amplitude but 
different phases \cite{coolswd}.
Fig.~\ref{cswd} depicts a sketch of a 
high fan-in majority gate in spin-wave technology.

In~\cite{fan2014_spin_mem}, a {\em Spin-Memeristor Threshold Logic} (SMTL) gate using memristive crossbar array is proposed. There, an array of SMTL gates is designed and simulated with experimentally validated device model characteristics. By varying the threshold input count, different possible mappings are demonstrated with good performance improvement over CMOS FPGA structures. 

A programmable CMOS/memristor threshold logic is proposed in~\cite{gao13_cmtl}. A $4$-input threshold logic gate is experimentally demonstrated using Ag/a-Si/Pt memristive devices. They also propose a threshold logic network similar to~\cite{fan2014_spin_mem} with programmable fan-in.

It is to be noted that none of the aforementioned implementations employed any automated synthesis flow to exploit majority gates with larger than $3$ inputs. Thus, the potential of compact realization of diverse applications, even if feasible with these technologies, is hardly experimented due to the lack of an efficient synthesis flow. Our proposed sound and complete axiomatization aims at filling this gap.

\vspace{0.02in}

Note that the aforementioned examples are just few of the possible applications of $n$-ary majority logic and of its sound and complete axiomatization. More opportunities exist in other fields of computer 
science
  but their discussion is out of the scope of this 
paper. 

\section{Conclusions}
\label{concl}

In this paper, we proposed a sound and complete axiomatization of majority logic. Stemming from previous work on MAJ-3/INV logic, we extended fundamental axioms to 
arbitrary $n$-ary majority operators. Based on this general set of axioms, computer applications 
can
now fully exploit the expressive power of majority logic. We discussed the potential impact in the fields of logic optimization, Boolean satisfiability, repetition codes
and emerging technologies.
From a general standpoint, the possibility of manipulating logic in terms of majority operators paves the way for
more efficient computer applications where the core reasoning tasks are performed in the Boolean domain. 
In particular, possible directions for future work include 
 the 
development of (i) a complete majority satisfiability solver
and (ii) a majority synthesis tool targeting nanotechnologies.

\section*{Acknowledgements}

The authors would like to thank Prof. Maciej Ciesielski for valuable discussions. 
This research was supported by ERC-2009-AdG-246810.

\end{document}